\documentclass[aps,notitlepage,twocolumn,showpacs,preprintnumbers,superscriptaddress,floatfix,tightenlines]{revtex4-1}
\usepackage[T1]{fontenc} 

\usepackage{soul}
\usepackage{mathrsfs}
\usepackage{amsmath}
\usepackage{amsfonts}
\usepackage{booktabs}
\usepackage{siunitx}
\usepackage{bbm}
\usepackage{array}
\newcolumntype{M}[1]{>{\centering\arraybackslash}m{#1}}
\usepackage{verbatim}
\usepackage{float}
\usepackage{xcolor}
\definecolor{lcolor}{rgb}{0.5,0,0}
\definecolor{citcolor}{rgb}{0,0.3,0.0}
\usepackage[breaklinks,colorlinks,urlcolor=blue,citecolor=citcolor,linkcolor=lcolor]{hyperref}

\def\bea{\begin{eqnarray}}
\def\eea{\end{eqnarray}}
\def\be{\begin{equation}}
\def\ee{\end{equation}}

\newcommand{\fig}{Fig.~}

\newcommand{\eq}{Eq.~}

\newcommand{\Tr}{\mathrm{Tr}}
\newcommand{\mbf}{\mathbf}

\newcommand{\pToFigs}{images}
\newcommand{\new}[1]{#1}

\usepackage{xargs}
\usepackage[colorinlistoftodos,prependcaption,textsize=footnotesize]{todonotes}
\newcommandx{\DM}[2][1=]{\todo[linecolor=orange,backgroundcolor=orange!25,bordercolor=orange,#1]{DM: #2}}
\newcommandx{\KB}[2][1=]{\todo[linecolor=purple,backgroundcolor=purple!25,bordercolor=purple,#1]{KB: #2}}
\newcommandx{\PH}[2][1=]{\todo[linecolor=yellow,backgroundcolor=yellow!25,bordercolor=yellow,#1]{PH: #2}}
\newcommandx{\TODO}[2][1=]{\todo[linecolor=red,backgroundcolor=red!25,bordercolor=red,#1]{TODO: #2}}

\begin{document}

\title{Real-time correlators in 3+1D thermal lattice gauge theory}

\author{Kirill Boguslavski}
\email{kirill.boguslavski@tuwien.ac.at}

\author{Paul Hotzy} 
\email{paul.hotzy@tuwien.ac.at}

\author{David I.~M\"uller} 
\email{dmueller@hep.itp.tuwien.ac.at}
\affiliation{Institute for Theoretical Physics, Technische Universit\"at Wien, 1040 Vienna, Austria}

\begin{abstract}
 We present the first direct ab-initio computation of unequal-time correlation functions in non-Abelian lattice gauge theory. We demonstrate non-trivial consistency relations among correlators, time-translation invariance, and agreement with Monte-Carlo results for thermal equilibrium in 3+1 dimensions by employing our stabilized complex Langevin method. Our work sets the stage to extract real-time observables, relevant to quark-gluon plasma physics within a first-principles real-time framework.
\end{abstract}

\maketitle



\section{Introduction}

The real-time dynamics of quantum fields in and out of equilibrium describe some of the most interesting and important phenomena of our universe ranging from cosmological to subatomic scales.
Theoretical predictions for real-time quantum dynamics are imperative for testing our understanding of such phenomena.
However, a description from first principles is typically either absent or poses great computational challenges for sufficiently complex systems.  

Of special interest to high-energy particle physics is the evolution of the strongly interacting medium consisting of quarks and gluons. Known as the quark-gluon plasma (QGP), it has likely existed in the earliest instants of our universe. On Earth, it is formed in relativistic heavy-ion collision experiments at large accelerator facilities such as RHIC and the LHC \cite{Busza:2018rrf}.
Our primary motivation in this work is to study the real-time dynamics of the QGP. It is described by quantum chromodynamics (QCD), one of the fundamental building blocks of the Standard Model of particle physics. 
The most successful method for making non-perturbative predictions for this theory is lattice QCD, which is usually restricted to real-time-independent observables \cite{FlavourLatticeAveragingGroupFLAG:2021npn}.
In the absence of an ab-initio approach for real-time QCD, one must rely on effective and phenomenological models. 
A particularly successful description of the QGP is provided by relativistic hydrodynamics \cite{Romatschke:2017ejr}. However, hydrodynamics requires input from the underlying theory in terms of viscosities. 
Moreover, effective descriptions of experimental probes like jets and heavy quarks also rely on the knowledge of QCD transport coefficients \cite{Apolinario:2022vzg}.

\begin{figure}[t]
    \centering
    \includegraphics{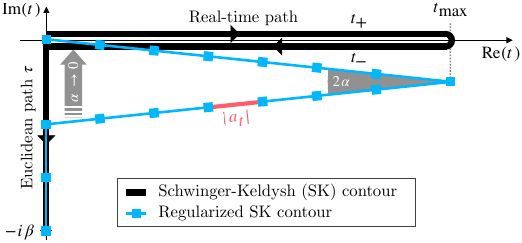}
    \caption{
    Schwinger-Keldysh (SK) contour (black) in the complex time plane and a regularized SK contour (blue) with tilt angle $\alpha$ and complex time step $a_t$.
    Both contours include a (tilted) real-time path and a Euclidean path. \new{The presented figure is a schematic sketch of the SK-contour and its lattice discretized version. The scales and number of points depicted do not reflect the actual model parameters.}
    \label{fig:sk_contour}
    }
\end{figure}

All of these real-time observables can be extracted from unequal-time correlation functions. In practical simulations, such quantities suffer from a numerical sign problem \cite{Gattringer:2016kco}, as we exemplify by the correlation of an arbitrary time-dependent observable $O(t)$. Expressed as a path integral over all realizations of the field $A$,
\begin{align}
    \langle  O(t) O(t') \rangle = \frac{1}{Z}\int \mathcal D A\, e^{i S[A]} O(t) O(t'),
    \label{eq:OO_pathint}
\end{align}
the correlator involves a complex-valued weight $\exp(iS)$. 
The highly oscillatory nature of this integral impedes the numerical application of standard Monte Carlo techniques. This becomes particularly hard in Minkowski space-time since the action $S[A]$ is real-valued.

Even though the sign problem has evaded a general and efficient solution due to its NP-hardness \cite{Troyer:2004ge}, progress on extracting observables can be made for individual systems nonetheless. Numerical simulations of real-time scalar fields have been carried out recently using
the functional renormalization group \cite{Pawlowski:2015mia, Huelsmann:2020xcy, Roth:2023wbp} and contour deformation \cite{Alexandru:2016gsd, Alexandru:2017lqr}. 
For QCD, transport coefficients and spectral functions have been computed using spectral reconstruction and analytic continuation in Euclidean lattice gauge theory \cite{Asakawa:2000tr, Meyer:2007dy, Burnier:2014ssa, Altenkort:2022yhb, Rothkopf:2022ctl, Altenkort:2023oms}, which forms an ill-posed inverse problem. Consequently, extracting accurate results becomes computationally challenging.
Interesting alternative approaches employ analog and digital quantum simulators of gauge theories \cite{Zohar:2015hwa, Martinez:2016yna, Banuls:2019bmf}, which are currently limited in their complexity and lattice size.

In this work, we employ a stabilized version of the complex Langevin (CL) method to evade the sign problem. We achieve the first stable ab-initio computation of gauge-invariant real-time correlation functions in thermal SU(2) lattice gauge theory in 3+1 dimensions. This represents a breakthrough that in the future may pave the way to extract transport coefficients and spectral functions directly.

This article is structured as follows: Section \ref{sec:cl} introduces the CL method applied to Yang-Mills theory on a real-time lattice. We discuss our simulation approach including our stabilization and extrapolation strategies. In Sec.\ \ref{sec:results} we show our numerical results for unequal-time correlation functions and conclude in Sec.\ \ref{sec:conclusion}. Details of our approach, including further checks of correctness and how to simulate longer physical times, can be found in the Appendices.


\section{Complex Langevin method} \label{sec:cl}
The CL approach is based on stochastic quantization \cite{Parisi:1980ys}. In this method, the path integral expression in \eq \eqref{eq:OO_pathint} is substituted by an average over a stochastic process for the fields. Although stochastic quantization was originally proposed for real-valued weights $\exp(-S)$, it was soon extended to the complex case, which is known as the CL method \cite{Parisi:1983mgm}. For \eq \eqref{eq:OO_pathint}, the stochastic process of gauge fields is described by
\begin{align}
    \frac{\partial}{\partial \theta}  A^a_\mu(x, \theta)= \left.i \frac{\delta S}{\delta A^a_\mu(x)}\right\rvert_\theta+ \eta^a_\mu(x, \theta), \label{eq:CL}
\end{align}
where $x$ denotes the space-time point, $\theta$ the fictitious Langevin time and $\eta^a_\mu(x, \theta)$ is a real-valued Gaussian random field. 
The stationary limit $\theta \rightarrow \infty$ of the process $A^a_\mu(x, \theta)$ can be used to approximate the path integral. While issues with stability and wrong convergence can occur, important conceptual improvements including convergence criteria \cite{Aarts:2009uq, Nagata:2016vkn, Kades:2021hir} and numerical achievements have reinvigorated the field more recently. Adaptive \cite{Aarts:2009dg, Flower:1986hv} and implicit \cite{Alvestad:2021hsi} solvers for the CL equation have been shown to lessen problems associated with unstable runaway trajectories. For gauge theories, convergence properties have been improved by gauge cooling \cite{Seiler:2012wz}, which exploits the gauge freedom of the CL process. In the realm of QCD at finite chemical potential, gauge cooling partially in combination with dynamical stabilization \cite{Attanasio:2018rtq} has led to advances in the computation of the equation of state \cite{Bongiovanni:2013nxa, Sexty:2013ica, Mollgaard:2013qra, Fodor:2015doa, Nagata:2015uga, Aarts:2016qrv, Kogut:2019qmi, Attanasio:2020spv, Attanasio:2022mjd}. Recently, there has been renewed interest in kernels \cite{Okamoto:1988ru, Okano:1991tz}, which are modifications of the CL equation that potentially improve convergence properties. Machine-learning-based kernels have been successfully applied to real-time scalar fields in up to 1+1 dimensions \cite{Alvestad:2022abf, Lampl:2023xpb, Alvestad:2023jgl} extending the results of earlier studies with non-stabilized CL and contour deformation technique \cite{Berges:2005yt, Berges:2006xc, Alexandru:2020wrj}.
Further successful applications of the CL method include quantum many-body, cold atom, and spin systems \mbox{\cite{Aarts:2011zn, Rammelmuller:2018hnk, Berger:2019odf, Heinen:2023wtt}}. 

However, so far, the application of the real-time CL method to non-Abelian gauge theories has been elusive \cite{Berges:2006xc, Berges:2007nr, Aarts:2017hqp}. This has changed with our recent revision of the CL equation for non-Abelian gauge theories \cite{Boguslavski:2022dee, Hotzy:2022zhu}, which enables unprecedentedly stable simulations on complex time contours. Here we use this new approach to extract real-time correlation functions in Yang-Mills theory.


\subsection{CL for real-time lattice gauge theories}

We consider real-time SU($N_c$) Yang-Mills theory in 3+1 dimensions in thermal equilibrium. Formally, this system can be described within the Schwinger-Keldysh (SK) formalism \cite{Schwinger:1960qe, Keldysh:1964ud}, which puts the theory on a complex time contour $\mathscr C$ shown as the black curve in \fig\ref{fig:sk_contour}. The real part of the contour (real-time path) describes the extent up to $t_{\mathrm{max}}$ in physical Minkowski time with a forward ($t_+$) and backward ($t_-$) time path. In contrast, the Euclidean path follows the imaginary time axis ($\tau$) whose extent corresponds to the inverse temperature $\beta = 1/T$. The SK contour enters the action, 
\begin{align} \label{eq:YM_action}
    S[A] = - \frac{1}{2 g^2} \int_{\mathscr C} \mathrm{d} t \int \mathrm{d}^3 x\, \Tr[F^{\mu \nu} F_{\mu \nu}],
\end{align}
with the field-strength tensor $F^{\mu \nu}$ and the coupling constant $g$. 
The contour is closed, which implies periodic boundary conditions for the field \mbox{$A(t{=}-i\beta) = A(t{=}0)$}. 

To simulate this system, we use a standard lattice gauge theory formulation that guarantees gauge invariance by construction \cite{Wilson:1974sk}: we discretize the gauge fields on a lattice of size $N_t \times N_s^3$ and introduce unitary link variables \mbox{$U_{\mu}(x) \simeq \exp{\left[i a^\mu t^a  A^a_{\mu}(x+\hat \mu/2\big)\right]}$} (no sum over $\mu$), where $\hat \mu$ is a unit vector, $a^\mu$ are the lattice spacings and $t^a$ are the generators of SU($N_c$). The lattice analogue of \eq\eqref{eq:YM_action} is given by the Wilson action
\begin{align}
    S_\mathrm{w} = \frac{1}{g^2} \sum_{x, \mu \neq \nu} \rho_{\mu\nu}(x) \mathrm{Tr} \left[U_{\mu\nu}(x) - \mathbbm{1}\right],
\end{align}
with \mbox{$U_{\mu\nu}(x) = U_{\mu}(x)\, U_{\nu}(x+\hat{\mu})\, U^{-1}_{\mu}(x+\hat{\nu})\, U^{-1}_{\nu}(x)$} and prefactors $\rho_{0i}(x) = -a_s / a_t(x)$, $\rho_{ij}(x) = a_t(x)/a_s$. The SK contour enters through the time-dependent temporal spacings $a_t(x)$, whereas the spatial lattice spacing $a_s$ is constant. In this work, dimensionful quantities are given in units of $a_s$.

The discretized CL equation corresponding to \eqref{eq:CL} reads
\begin{align}
    U_{\mu}(x, \!\theta  \! + \!  \epsilon) = e^{i t^a \! \left[ i \epsilon\Gamma_{\!\mu}\!(x) \left.\frac{\delta S_\mathrm{w}}{\delta A^a_{\! \mu}\!(x)}\right\vert_\theta + \sqrt{ \epsilon\Gamma_{\!\mu}\!(x)}\, \eta^a_{\mu}\!(x, \theta) \right]} U_{\mu}(x,\! \theta),\label{eq:CL_lattice}
\end{align}
where $\epsilon$ represents the Langevin time step and $\Gamma_{\mu} (x)$ is the kernel. The Gaussian noise field satisfies
\begin{align}
    \langle \eta^a_\mu(x, \theta) \rangle = 0, ~\,
    \langle \eta^a_\mu(x, \theta) \eta^b_\nu(y, \theta') \rangle = 2 \delta_{\theta\theta'} \delta_{xy} \delta^{ab} \delta_{\mu\nu}.
\end{align}
Note that the real-time part of the SK contour leads to a complex-valued drift term $i \delta S_\mathrm{w} / \delta A^a_\mu(x)$. This necessitates the generalization of gauge links from SU($N_c$) to \mbox{SL($N_c$, $\mathbb{C}$)} and the analytical continuation of the action.

\subsection{Simulation and stabilization strategy} \label{sec:sim_strat}

In the CL approach, we have the freedom to introduce a kernel $\Gamma_\mu$ \cite{Okamoto:1988ru, Okano:1991tz}. Such a modification of the CL equation leaves the stationary solution unchanged. In our simulations, we employ
\begin{align} \label{eq:kernel}
    \Gamma_{0}(x) = |a_t(x)|^2 / a_s^2,
    \quad \Gamma_{i}(x) = 1.
\end{align}
This is motivated by a time-contour parametrization applied to the CL equation and originally introduced by us in \cite{Boguslavski:2022dee}. There we have demonstrated that in combination with gauge cooling, this kernel enhances the stability and convergence of our simulations systematically as the lattice anisotropy $a_s/\vert a_t \vert$ increases. 
\new{We note that field-independent kernels have been used in Euclidean Langevin simulations to shorten the autocorrelation time \mbox{\cite{Batrouni:1985jn}}, which, as we emphasize, has a different objective than in our case.}
Additionally, we employ an improved update step  \cite{Ukawa:1985hr} to mitigate systematic errors (see App.~\ref{app:impr_step}). 

In our simulations, we iteratively solve the discretized CL equation. At sufficiently late Langevin times, the gauge links are distributed according to the desired stationary probability density. We ensure this by computing observables such as Wilson loops and comparing them to Euclidean results where applicable. Expectation values $\langle {O} \rangle$ are calculated by sampling uncorrelated gauge configurations $\{U^{(i)}\}$
\begin{align}
    \langle {O} \rangle \approx \frac{1}{N_\mathrm{cfgs}} \sum_i^{N_\mathrm{cfgs}} {O}\big[U^{(i)}\big].
\end{align}
To further validate our simulations, we calculate the unitarity norm in App.~\ref{sec:un}. We find that it assumes small, stable values, which have been empirically associated with the correct convergence of CL \cite{Seiler:2012wz}.

We regulate the path integral \eqref{eq:OO_pathint} by introducing a tilt angle $\alpha > 0$ for the real-time part of the contour, as depicted in \fig \ref{fig:sk_contour}. This angle additionally softens the sign problem. While the discretized path integral for \mbox{$\alpha = 0$} is ill-defined \cite{Matsumoto:2022ccq}, the SK contour is reached in the limit $\alpha \rightarrow 0^+$. In our approach, we generate configurations for multiple tilt angles, compute expectation values $\langle  O \rangle_\alpha$ and obtain real-time observables in the $\alpha \rightarrow 0$ limit,
\begin{align}
    \langle  O \rangle = \lim_{\alpha \rightarrow 0^+} \langle  O \rangle_\alpha.
\end{align}
We illustrate this extrapolation in \fig\ref{fig:sk_contour} where the grey arrow symbolizes the convergence of the tilted regularized contour (blue) toward the SK contour (black). Details of this procedure are discussed in the following.


\subsection{Time contour discretization and extrapolation}

The discretized contour is shown in blue in \fig\ref{fig:sk_contour} 
and consists of two tilted real-time paths and an Euclidean path. It is defined by the real-time extent $t_\mathrm{max}$, its extent in imaginary time $\beta$, and the tilt angle $\alpha \in [0,\pi/2)$. We discretize the contour by choosing $N_\mathrm{tilt}$ points on each of the two tilted paths and $N_\mathrm{Euclid}$ points on the Euclidean path, such that the total number of temporal points is $N_t = 2 N_\mathrm{tilt} + N_\mathrm{Euclid}$. The complex temporal steps are
\begin{align}
    a_{t,k} = \begin{cases}
  +\tilde a  e^{-i \alpha} & 0 \leq k < N_\mathrm{tilt} \\
  -\tilde a  e^{+i \alpha} & N_\mathrm{tilt} \leq k < 2N_\mathrm{tilt} \\
  -i a_\tau  & 2N_\mathrm{tilt} \leq k < N_t
\end{cases},
\end{align}\new{where $k \in \{ 0,1,2, \dots, N_t-1 \}$ enumerates the points on the contour. We set
$\tilde a = t_\mathrm{max} / (N_\mathrm{tilt} \cos \alpha)$ and $a_\tau = (\beta - 2 t_\mathrm{max} \tan \alpha) / N_\mathrm{Euclid}$. 
Given specific values of the angle $\alpha$, we choose $N_\mathrm{tilt}$ and $N_\mathrm{Euclid}$ such that $\tilde a \approx a_\tau$ and thus $|a_{t, k}| \approx const$. In Table \mbox{\ref{tab:contour_params}} we provide the parameters used in our simulations. We denote the varying lattice spacing $a_{t,k}$ by $a_t(x)$. Here, the dependence on the lattice site $x$ is understood to reduce to the dependence on the contour point index $k$.
}

\begin{table}[tb]
    \caption{Numerical parameters of the discretized time contour in our simulations.}
    \label{tab:contour_params}
    \centering
    \begin{ruledtabular}
        \begin{tabular}{c|ccccc}
        $\tan \alpha$  & $N_\mathrm{tilt}$ & $N_\mathrm{Euclid}$ & $N_t$ & $\tilde a$ & $a_\tau$ \\
        \hline
        Euclidean & --  & 16 & 16 & \makebox[0pt][l]{\quad--}\phantom{0.0000} & 0.0625 \\
        $1/3$ & 25  & -- & 50 & 0.0633 & \makebox[0pt][l]{\quad--}\phantom{0.0000} \\
        $1/6$ & 24  & 8 & 56 & 0.0634 & 0.0625 \\
        $1/12$ & 24  & 12 & 60 & 0.0627  & 0.0625 \\
        $1/24$ & 24  & 14 & 62 & 0.0625 & 0.0625 \\
        $1/48$ & 24  & 16 & 64 & 0.0625 & 0.0586 \\
        $1/96$ & 24  & 16 & 64 & 0.0625 & 0.0606 \\
        \end{tabular}
    \end{ruledtabular}
\end{table}

Care has to be taken when observables are extrapolated to the SK contour ($\alpha \rightarrow 0$). First, we measure a time-dependent observable $\langle O_k \rangle_\alpha$ along the contour for different values of $\alpha$ in our simulations. Note that for each $\alpha$ the number of points $N^{(\alpha)}_\mathrm{tilt}$ may be different. Next, we resample the observable onto the finest discretization $N^*_\mathrm{tilt} = \max_\alpha N^{(\alpha)}_\mathrm{tilt}$ via polynomial interpolation. Using the resampled data for $\langle O_k \rangle_\alpha$, we then perform a cubic polynomial fit for the dependence on $\alpha$ for each contour grid point $k$. Finally, the cubic polynomials are extrapolated to $\alpha \rightarrow 0$ to obtain our final result. The same method can also be applied to arbitrary $n$-point correlation functions. We emphasize that the tilt angles need to be sufficiently close to zero in order for the extrapolation to be well-behaved.


\subsection{Numerical setup} \label{sec:setup}

We simulate SU(2) gauge theory in thermal equilibrium on a lattice with $N_s^3 = 16^3$ spatial lattice points. The number of temporal lattice points $N_t$ is chosen to maintain a constant anisotropy $a_s / \vert a_{t} \vert \approx 16$ along the entire regularized complex time contour (\fig \ref{fig:sk_contour}), while decreasing the tilt angle $\tan(\alpha) = 1/3 \rightarrow 1/96$. We employ the bare coupling $g=0.5$, the inverse temperature $\beta=1/T=1$, and a maximal real-time extent of $t_\mathrm{max} = 1.5 \beta$. We note that these parameter choices correspond to the deconfined regime: at $g = 0.5$, the Polyakov loop admits an expectation value of $\langle P \rangle \approx 0.98$ and the phase transition occurs roughly at $g_\mathrm{crit} \approx 2$. Our calculations are thereby conducted in a weakly coupled regime away from the true continuum limit. 

However, following indications of our previous study on complex time contours \cite{Boguslavski:2022dee}, we note that both $t_\mathrm{max}$ and $g$ can be in principle increased systematically by using a finer temporal discretization, albeit at a higher computational cost.
For instance, increasing the anisotropy to $a_s / \vert a_{t} \vert = 128$, and thus also $N_t$, enables us to obtain correct results for $t_\mathrm{max} = 2\beta$, as showcased in App.\ \ref{app:2beta}.

In addition to the kernel in \eq \eqref{eq:kernel}, we apply gauge cooling \cite{Seiler:2012wz} with one cooling step after each update, using a step size $\alpha_{\mathrm{GC}}=0.05$, to stabilize our simulations (see \cite{Boguslavski:2022dee} for details). 
The simulations start cold with unit matrices $U_{x,\mu}(\theta \! = \!0) = \mathbbm{1}$ and evolve with a constant Langevin time step $\epsilon=10^{-4}$. Field configurations for the measurement of observables are extracted within $10 \leq \theta \leq 20$ after thermalization at Langevin times separated by $\Delta \theta = 0.1$, which is well above the auto-correlation time of the observables we are interested in. 
We also average over configurations generated from 100 to 1000 independent simulations. Error bars in the presented figures are determined using a bias-corrected jackknife method.


\section{Unequal real-time correlations}
\label{sec:results}

An important class of observables accessible in real-time simulations of lattice gauge theory are unequal time correlation functions of the energy-momentum tensor
\begin{align} \label{eq:tmunu_corr}
    C_{\mu\nu; \rho\sigma}(t, \mbf x; t', \mbf x') = \langle T_{\mu\nu}(t, \mbf x) T_{\rho\sigma}(t', \mbf x') \rangle,
\end{align}
with $T_{\mu\nu} = 2 \Tr [F_{\mu}{}^\alpha F_{\alpha\nu} + \frac{1}{4} g_{\mu\nu}F_{\alpha\beta} F^{\alpha\beta}]$.
The spectral function associated with \eq\eqref{eq:tmunu_corr} contains information about the transport properties of the system. In particular, $C_{\mu\mu;\nu\nu}$ and $C_{xy;xy}$  encode the bulk and shear viscosities $\eta$ and  $\zeta$ entering hydrodynamic equations \cite{Romatschke:2017ejr}.

In this work, we focus on the magnetic contribution to the energy density calculated in terms of cloverleaves (see App.\ \ref{app:clover})
\begin{align}
    O(t, \mbf x) = 
    \frac{1}{2} \Tr[ F_{ij}(t, \mbf x) F^{ij}(t, \mbf x)],
    \label{eq:O_def}
\end{align}
and the corresponding unequal-time correlator
\begin{align} \label{eq:corr_fun}
    \! \! \! C(t, \! \mbf x; t', \!  \mbf x') \!&=\! \langle O(t, \! \mbf x) O(t', \! \mbf x')  \rangle \! -\!   \langle O(t, \! \mbf x)  \rangle \langle O(t', \! \mbf x')  \rangle, \\
    C(t,t') &= \frac{1}{ N_s^3}\sum_{\mbf x} C(t, \mbf x; t', \mbf x).
\end{align}
The main advantage of studying $C(t,t')$ is its close relation to the correlator of the energy-momentum tensor \eqref{eq:tmunu_corr}, while requiring fewer configurations due to the average over the spatial lattice sites $\mbf x$. Nonetheless, $C(t,t')$ exhibits non-trivial features that are manifest to such correlators in the SK limit $\alpha \rightarrow 0^+$, many of which we will explicitly check numerically.

\subsection{Statistical correlation and spectral function}

\begin{figure}
    \centering
    \includegraphics[width=\linewidth]{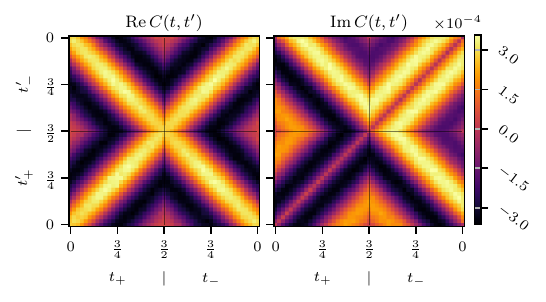}
    \caption{Real part ({\em left}) and imaginary part ({\em right}) of the unequal-time correlation function $C(t,t')$ extrapolated onto the SK contour using a cubic polynomial in $\alpha$. Forward and backward paths are indicated by $t_\pm$. 
    \label{fig:corr_contour_plot}
    }
\end{figure}

We present our main result in \fig\ref{fig:corr_contour_plot}. It shows the correlation function $C(t,t')$ extrapolated to $\alpha \rightarrow 0^+$ and restricted to the real-time forward and backward paths. A striking feature of \fig\ref{fig:corr_contour_plot} is that $C(t,t')$ splits into four distinct quadrants, where each quadrant represents a propagator, 
\begin{equation}
\begin{aligned}
    D^<(t,t') = C(t_+, t_-'), & &D^{\bar F}(t,t') = C(t_-, t_-'),\\
    D^F(t,t') = C(t_+, t_+'), & &D^>(t,t') = C(t_-, t_+'),
\end{aligned}
\end{equation}
and $t$ is either $t_+$ or $t_-$, and similarly for $t'$. Here $D^F$ and $D^{\bar{F}}$ are known as (anti-)Feynman propagators, and $D^>$ and $D^<$ are Wightman functions. Additionally, we see that $C(t,t')$ exhibits a symmetry: in each quadrant, we find that the propagators become independent of the central time coordinate $\bar t = (t+t')/2$:
\begin{align}
    D(t,t') = D(t-t') \equiv D(\Delta t).
\end{align}
This symmetric structure of $C(t,t')$ is a direct consequence of time translation invariance in thermal equilibrium and only appears in the SK limit $\alpha \rightarrow 0^+$, as we will show below.

\begin{figure}[tb]
    \centering
    \includegraphics{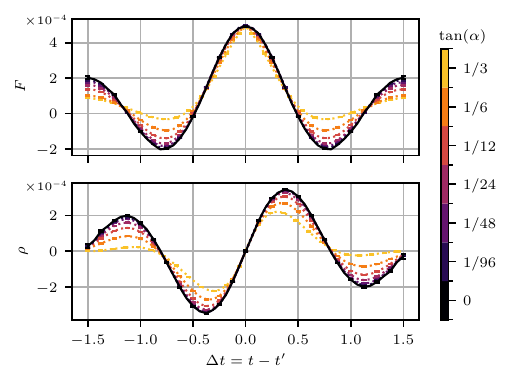}
    \caption{
    Statistical correlation function $F$ and spectral function $\rho$ as functions of the real-time difference $\Delta t = t - t'$ for varying tilt angles $\alpha$. 
    For finite $\alpha$, the real times $t, t'$ are obtained by projecting the tilted time contour onto the real-time axis. 
    The data extrapolates toward $\alpha \to 0$, indicating a clear convergence to finite real-time observables.
    \label{fig:limit} }
\end{figure}

In \fig\ref{fig:limit}, we present results for the statistical correlation function and the spectral function
\begin{align}
    F = \mathrm{Re}\, D^F, ~~
    \rho = -\mathrm{sgn}(\Delta t)\,\mathrm{Im}\, D^F,
\end{align}
averaged over the central time $\bar t$ at various tilt angles of the discretized time contour, with signum function $\mathrm{sgn}(\Delta t)$. 
As the tilt angle decreases, both $F$ and $\rho$ converge to well-defined curves (black lines), demonstrating successful extrapolation.

\subsection{Euclidean correlator}

\begin{figure}[tb]
    \centering \includegraphics{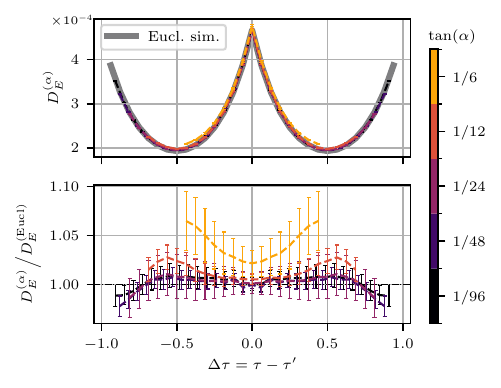}
    \caption{
        \new{Euclidean correlator of the magnetic contribution to the energy density.
         (\emph{top})  Euclidean correlator $D_E^{(\alpha)}$ at different tilt angles. (\emph{bottom}) Ratio $D_E^{(\alpha)}/D^\mathrm{(Eucl)}_E$ where $D^\mathrm{(Eucl)}_E$ is extracted from a Euclidean simulation. The ratio approaches one in the limit $\alpha \rightarrow 0$. The small deviations stem from statistical errors and minor differences in the lattice spacing.}
    }
    \label{fig:euclidean_corr}
\end{figure}

We simulate the gauge fields on the discretized SK contour -- this gives us also access to the Euclidean correlator
\begin{align}
    D_E(\tau, \tau') = C(\tau, \tau'),
\end{align}
where $\tau$ and $\tau'$ are restricted to the imaginary Euclidean path of the contour (see \fig \ref{fig:sk_contour}). Our results are shown in \fig\ref{fig:euclidean_corr}, where we present $D_E$ for various values of the tilt angle $\alpha$ and compare these correlators to the result of a Euclidean simulation (grey), where no sign problem is present. \new{The Euclidean results are obtained by real Langevin simulations \mbox{\cite{Parisi:1980ys}}.} We find remarkable agreement for a wide range of $\alpha$, showing the consistency of our simulations on the SK contour. We emphasize that due to the non-locality in Euclidean time, this result is significantly more robust to indicate correct convergence than the comparisons conducted in \cite{Berges:2006xc, Boguslavski:2022dee} where only time-translation invariant one-point functions have been used. \new{This indicates the absence of boundary terms, which suggests that the criterion for correct convergence is fulfilled \mbox{\cite{Aarts:2009uq, Scherzer:2018hid, Seiler:2023kes}}.}

\subsection{Emergent consistency among propagators}

\begin{figure}[t]
    \centering
    \includegraphics{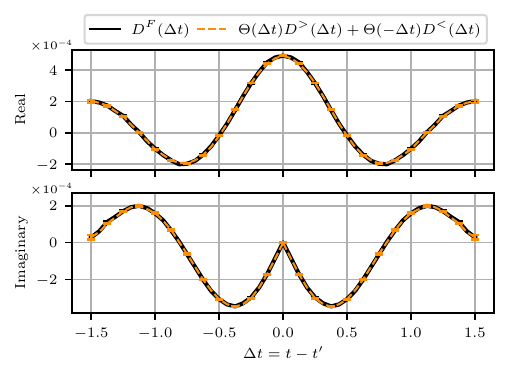}
    \caption{Relation between Feynman propagator and Wightman functions extracted independently from different quadrants of $C(t,t')$ in \fig \ref{fig:corr_contour_plot}. The top and bottom panels show real and imaginary parts of \eq\eqref{eq:feyn_vs_wight}, respectively. Our simulations demonstrate that this consistency relation is satisfied with remarkable accuracy.
    \label{fig:relation_corrs}
    }
    \label{fig:wightman_vs_feynman}
\end{figure}

There are well-established relations in field theory for different correlation functions. Analytically, the Feynman propagator can be expressed in terms of Wightman functions \cite{Ghiglieri:2020dpq}
\begin{align} \label{eq:feyn_vs_wight}
    D^F(t, t') = \Theta(t\! -\! t') D^>(t,t') \! +\!  \Theta(t'\! -\! t) D^<(t, t'),
\end{align}
where $\Theta(t{-}t')$ is the Heaviside step function. With our approach, we can reproduce this correspondence numerically. In \fig \ref{fig:relation_corrs} we show that in the limit $\alpha \to 0^+$, \eq\eqref{eq:feyn_vs_wight} is indeed satisfied. 

\begin{figure}[t]
    \centering
    \includegraphics{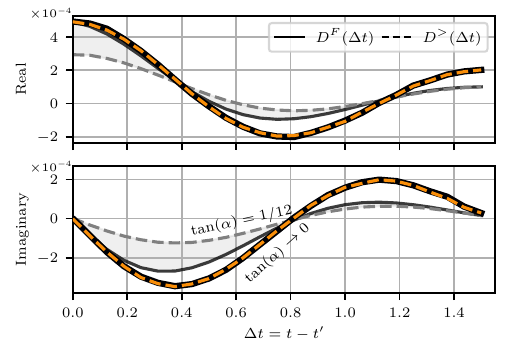}
    \caption{
    Explicit check of \eq \eqref{eq:feyn_vs_wight} for $t'>t$ with vanishing tilt angle $\alpha\to 0$ (black and orange) and finite $\tan(\alpha)=1/12$ (grey). One observes that the non-trivial correspondence \eqref{eq:feyn_vs_wight} between these correlations only emerges for $\alpha\to 0$.
    \label{fig:corr_rel}}
\end{figure}

In contrast, \fig \ref{fig:corr_rel} shows that the relation \eqref{eq:feyn_vs_wight} is not satisfied for finite tilt angle such as $\tan(\alpha)=1/12$, as indicated by the grey-shaded region that highlights the deviation between both correlators. It rather emerges in the limit $\alpha\to0^+$, where the relation is satisfied to high accuracy. This underpins the necessity for the extrapolation procedure of our simulation strategy. The numerical consistency among different correlation functions represents a non-trivial property manifest in the real-time evolution. 

\subsection{Emergent time translation invariance} \label{sec:transl_inv}

In thermal equilibrium, observables are time translation invariant. For one-point functions, this implies $\langle {O}(t) \rangle = \mathrm{const}$ while two-point functions $\langle {O}(t) {O}(t') \rangle$ are independent of the central time $(t+t')/2$ and only depend on the time difference $\Delta t \equiv t-t'$. 
However, we find that this time translation invariance is violated for the regularized SK contour. The reason for this is as follows: at finite angles $\alpha$, real-time values are extracted by projecting $t$ and $t'$ onto the real-time axis. The tilt angle pulls apart different points on the forward and backward branches of the SK contour for the same real-time. This can effectively introduce an unphysical dependence on the central time for correlation functions, hence violating time translation invariance. This symmetry is restored in the limit of vanishing tilt angle.

\begin{figure}[t]
    \centering
    \includegraphics{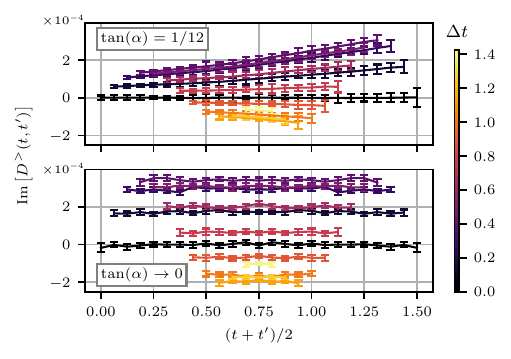}
    \caption{The imaginary part of the Wightman function $D^>$ is shown as a function of central time $(t+t')/2$ for different fixed $\Delta t = t-t'$ slices, indicated by different colors. 
    The figure shows the correlation for a finite tilt angle $\tan(\alpha)=1/12$ (\emph{top}) and the extrapolated correlation for $\tan(\alpha)\to 0$ (\emph{bottom}).
    We stress that the time translation invariance is broken for finite tilt angles while it is satisfied for the extrapolated data.
    \label{fig:t_indep}}
\end{figure}

In \fig\ref{fig:t_indep}, we numerically confirm our expectations for the imaginary part of the Wightman function $D^>$. It shows the correlation function as a function of the central time $(t+t')/2$ for varying time differences $\Delta t$, each represented by a different color. Horizontal lines in this representation indicate that the time translation invariance is intact. We observe that this feature is not present at the finite tilt angle $\tan(\alpha)=1/12$ shown in the top panel. When we extrapolate this correlator to the SK contour $\alpha \to 0$, we find that the independence of the central time becomes well-preserved, as depicted in the bottom panel. A similar assertion holds for other correlation functions that we have calculated in this study; however, it is most pronounced in the case of the Wightman functions as they reflect correlations between forward and backward real-time branches. Additionally, we emphasize that not only time independence is violated but also the values of the correlations are systematically distorted with respect to the central time.

\section{Conclusion} \label{sec:conclusion}
We have performed the first direct computation of unequal-time correlation functions in 3+1 dimensional real-time Yang-Mills theory in thermal equilibrium. These results are achieved by utilizing the complex Langevin (CL) method that we revised for complex time contours in \cite{Boguslavski:2022dee} and applied here to the Schwinger-Keldysh (SK) contour using a polynomial extrapolation. 

We have found that our new setup allows us to reliably extract real-time observables, as demonstrated by the correlation function of the magnetic contribution to the energy density. An important result is that the extracted correlators on the SK contour satisfy numerically non-trivial consistency relations that connect Wightman and Feynman propagators and are time-translation invariant. In contrast, such properties are violated on other complex time contours. Moreover, we have verified that our Euclidean correlator along the thermal path of the SK contour agrees with independent simulations using a traditional Monte Carlo method. 

These results give a strong indication that our method can be extended to other gauge-invariant observables such as correlations of the energy-momentum tensor $T_{\mu\nu}$ or other transport coefficients for heavy quarks and jets. \new{While the generalization to SU(3) is straightforward, stable simulations at larger couplings $g$ and real-time extents $t_\mathrm{max}$ require significantly more computational resources with the currently available methods. Therefore, applications may need further stabilization strategies in practice. We emphasize that the simplicity of our approach allows the combination with other possible stabilization techniques such as field-dependent kernels. With additional improvements, CL simulations could be used to access the spectral functions of various operators.} So far, their direct non-perturbative real-time computation in gauge theories can be performed in classical-statistical lattice simulations \cite{Boguslavski:2018beu, Boguslavski:2020tqz, Boguslavski:2021buh, Boguslavski:2021kdd, Ipp:2020mjc, Ipp:2020nfu, Avramescu:2023qvv}, which are justified far from equilibrium at weak couplings and large occupancies. To avoid these underlying approximations, another exciting prospect of our framework is the simulation of gauge theories outside of thermal equilibrium. This would improve our understanding of the thermalization process of gauge theories, which has significant phenomenological consequences, particularly in the context of the pre-equilibrium phase of heavy-ion collisions.


\vspace{1em}
\begin{acknowledgments}
We would like to thank J.~Berges, D.~Sexty, J.~Pawlowski, R.~Pisarski and A.~Rebhan for valuable discussions. The authors acknowledge funding from the Austrian Science Fund (FWF) projects \mbox{P~34455} and \mbox{W~1252}. D.M.~acknowledges additional support from project P~34764. The computational results presented here have been achieved using the Vienna Scientific Cluster (VSC).
\end{acknowledgments}

\appendix

\section{Improved Langevin step} \label{app:impr_step}
In our CL simulations, we utilize an improved update scheme that replaces \eq\eqref{eq:CL_lattice} in Sec.~\ref{sec:cl}. This improves the convergence of the algorithm and alleviates systematic errors stemming from a finite Langevin step size \cite{Ukawa:1985hr, Fukugita:1986tg}. 
The update step reads
\begin{align}
    \label{eq:improved_step}
    U_{\mu}(x, \theta + \epsilon) = e^{i \gamma t^a \left[ i \epsilon\Gamma_{\mu} \widetilde{\frac{\delta S_\mathrm{w}}{\delta A^a_{\mu}}} + \sqrt{ \epsilon\Gamma_{\mu}}\, \eta^a_{\mu}(x, \theta) \right]} U_{\mu}(x,\theta), 
\end{align}
with $\gamma = 1 + \epsilon \,C_A / 6$ and the quadratic Casimir $C_A=2$ for 
$\mathrm{SU}(2)$. 
The effective drift term is given by 
\begin{align}
    \widetilde{\frac{\delta S_\mathrm{w}}{\delta A^a_{\mu}}} = \frac{1}{2} \left(\frac{\delta S_\mathrm{w}}{\delta A^a_{\mu}}\Big[U\Big] + \frac{\delta S_\mathrm{w}}{\delta A^a_{\mu}}\Big[\widetilde{U}\Big]\right),
\end{align}
and averages the usual drift at Langevin time $\theta$ (denoted by links $U$) with an auxiliary drift with links
\begin{align}
    \widetilde{U}_{\mu}(x,\theta\!+\!\epsilon) = e^{i t^a \left[ i \epsilon\Gamma_{\mu} \frac{\delta S_\mathrm{w}}{\delta A^a_{\mu}}[U] + \sqrt{ \epsilon\Gamma_{\mu}}\, \eta^a_{\mu}(x, \theta) \right]} U_{\mu}(x,\theta).
\end{align}
The noise correlator is altered to
\begin{align}
    \langle \eta^a_{\mu}(x, \theta) \eta^b_{\nu}(y, \theta') \rangle = 2 (1 - \epsilon\, C_A/2)\, \delta_{\theta \theta'} \delta_{xy} \delta^{ab} \delta_{\mu\nu}.
\end{align}
This procedure may be understood as a second-order Runge-Kutta-Munthe-Kaas solver for differentiable Lie groups \cite{munthe1998runge}, adapted for stochastic differential equations. 

\begin{figure}[t]
    \centering
    \includegraphics{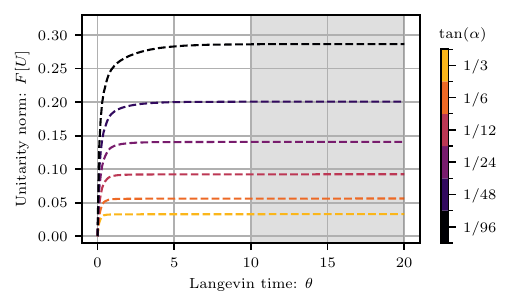}
    \caption{Unitarity norm $F[U]$ in \eqref{eq:unitarity_norm} as a function of the Langevin time $\theta$ for different tilt angles of the simulated tilted time contour.  As the tilt angle decreases, higher levels of non-unitarity are observed. Crucially, after a quick initial growth, all simulations lead to a plateau of the unitarity norm. The area highlighted by the grey background indicates where observables are measured. \label{fig:unitarity_norm}
    }
\end{figure}

\begin{figure}[t]
    \centering
    \includegraphics{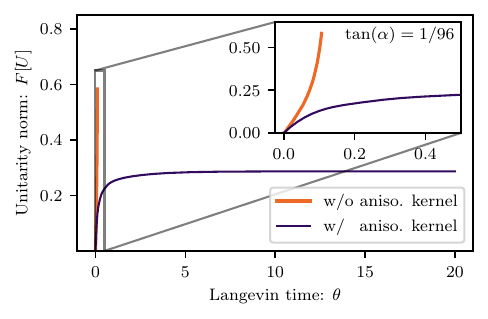}
    \caption{\new{Unitarity norm $F[U]$ as a function of the Langevin time $\theta$ with (dark purple line) and without (orange) the use of our anisotropic kernel. Simulations without the kernel exhibit fast growth and become unstable after a short time. }  \label{fig:comparision_un}
    }
\end{figure}

\section{Stability and the unitarity norm} \label{sec:un}
Empirical observations have shown that the stability of the CL evolution is closely associated with the ``non-unitarity'' of the field configuration \cite{Seiler:2012wz}. This property is typically quantified by the unitarity norm
\begin{align}
    \label{eq:unitarity_norm}
    F[U] = \frac{1}{4 N_t N_s^3}\sum_{x,\mu} \Tr\left[ (U_{\mu}(x) U_{\mu}^\dagger(x) - \mathbbm{1})^2 \right],
\end{align}
where $U_{\mu}(x)\in\mathrm{SL(N_c, \mathbb{C})}$ denote the link variables. Unitary link configurations have a vanishing unitarity norm $F[U]=0$.

In all simulations, we initialize the field with identity matrices $U_\mu(x,\theta \!\! = \!\!0) = \mathbbm{1}$, thereby starting at $F[U]=0$. Figure \ref{fig:unitarity_norm} shows the unitarity norm with respect to Langevin time, averaged over all runs that we used to evaluate the correlation function.
We observe that reducing the tilt angle leads to an increase in $F[U]$, indicating a more challenging sign problem. Crucially, however, the unitarity norm reaches a plateau after the thermalization of the CL process for all tilt angles. This suggests that the application of the anisotropic kernel in conjunction with the gauge cooling procedure successfully stabilizes the simulations. Observables are measured when the plateau is reached, and thermalization can be assumed. This region, $10 \leq \theta \leq 20$, is highlighted by the grey band in \fig \ref{fig:unitarity_norm}.

\new{Simulations without the use of the anisotropic kernel quickly converge to wrong results, or in the case of small tilt angles, even become unstable. The latter is demonstrated in \fig\mbox{\ref{fig:comparision_un}}, where we compare a \emph{kerneled} and an \emph{unkerneled} CL process for $\tan(\alpha) = 1/96$, leaving all other numerical and physical parameters the same.}

\begin{figure}[t]
    \centering\includegraphics{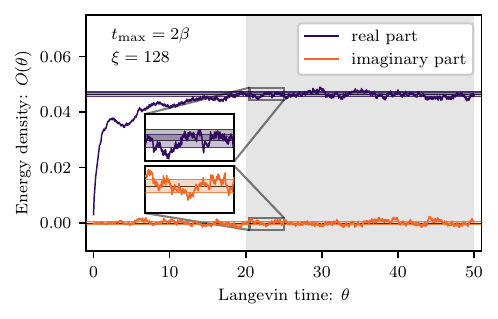}
    \caption{\new{Real (purple) and imaginary (orange) part of the magnetic contribution to the energy density $O$ as a function of the Langevin time $\theta$ for the larger real-time extent of $t_{\mathrm{max}}=2\beta$ and tilt angle $\tan(\alpha)=1/48$ at the same coupling $g=0.5$. To obtain a sufficient degree of stability, a lattice anisotropy of $\xi=128$ was employed. The grey-shaded background indicates the sampling region of the stochastic process for determining the expectation values, which are indicated by blue and orange bands for the real and imaginary parts. The correct value, obtained by Euclidean simulations, is shown as a black error band. The CL simulations show good agreement with the correct result.}
    \label{fig:larger_tmax}
    }
\end{figure}

\section{Simulating longer physical times} \label{app:2beta}

In our previous work \mbox{\cite{Boguslavski:2022dee}}, we established that the anisotropic kernel counteracts the increasing instabilities encountered for shrinking tilt angles by tweaking the anisotropy. Analogously, we can extend the simulated real-time at the same tilt angle by increasing the anisotropy. In \mbox{\fig \ref{fig:larger_tmax}}, we showcase this effect by using the magnetic contribution to the energy density of \eq \eqref{eq:O_def} averaged over time and space $O \equiv \frac{1}{N_t N_s^3}\sum_{t, {\mbf x}} O(t, \mbf x)$ for a system with a larger maximum real-time extent of $t_{\mathrm{max}}=2\beta$ as compared to the system with $t_{\mathrm{max}}=1.5\beta$ studied in the main text. We used the lattice anisotropy $\xi \equiv a_s/|a_t| = 128$  to stabilize the simulations since $\xi = 16$, as in the main text, is insufficient. The blue and orange curves show the CL trajectories of the real and imaginary parts of the magnetic contribution of the energy density, respectively. The black error band indicates the correct expectation value with the statistical error computed via Euclidean Langevin simulations. The expectation values of the CL simulation, computed from the sampling range highlighted in grey, are shown as bands in the respective colors for the real and imaginary parts. The overlap of the bands shown in the main panel and the insets suggests that the CL simulation successfully reproduces the correct expectation value. This confirms that the anisotropic kernel can be utilized to perform CL simulations for longer physical times.

\section{Cloverleaf definition of the magnetic energy density} \label{app:clover}

On the lattice, we obtain the field-strength tensor by relating it to the plaquette variable
\begin{align}
    U_{\mu\nu}(x) = \exp\left[ia_\mu a_\nu F_{\mu\nu}(x) + \mathcal{O}(a^3) \right].
\end{align}
This allows us to determine the magnetic contribution to the energy density as
\begin{align}
    O(t, \mbf x) &= \frac{1}{2}\, \Tr[ F_{ij}(t, \mbf x) F^{ij}(t, \mbf x)] \nonumber \\
    &\approx - \sum_{i<j} \frac{1}{a_i^2 a_j^2}\, \mathrm{Tr}\left\{ \mathcal{P}_A( C_{ij}(x) )^2\right\},
\end{align}
with 
\begin{align}
    \mathcal{P}_A(C)\equiv \frac{1}{2} \left( C- C^{-1} - \frac{1}{N_c} \Tr\left(C - C^{-1}\right)\right).
\end{align}
For an $\mathrm{SU}(N_c)$ matrix, this expression reduces to the anti-hermitian trace-zero part of $C$. 
The cloverleaf $C_{\mu\nu}(x)$ is given by
\begin{align}
\begin{split}
    C_{\mu\nu}(x) = \frac{1}{4} [&U_{\mu\nu}(x) + U_{\nu(-\mu)}(x) +   \\
     &U_{(-\mu)(-\nu)}(x) + U_{(-\nu)\mu}(x)]
\end{split}
\end{align}
and forms an average of four neighboring plaquettes. In contrast to using the plaquettes themselves, this results in a quantity that is defined on the lattice site $x$ and reduces lattice artifacts \cite{Bilson-Thompson:2002xlt}. 


\bibliography{main.bib}

\end{document}